\documentclass[english]{llncs}
\usepackage[T1]{fontenc}
\usepackage[latin9]{inputenc}
\usepackage{graphicx}
\usepackage{babel}
\usepackage{pbox}
\usepackage{caption,subcaption}
\begin{document}

\title{A Roofline Visualization Framework}

\author{Wyatt Spear and Boyana Norris}

\institute{University of Oregon}
\maketitle
\begin{abstract}
The Roofline Model and its derivatives provide an intuitive representation of 
the best achievable performance on a given architecture. 
The Roofline Toolkit project is a collaboration among researchers at 
Argonne National Laboratory, Lawrence Berkeley National Laboratory, and the University of 
Oregon and consists of three main parts: hardware characterization, software characterization, 
and data manipulation and visualization interface. These components address the different 
aspects of performance data acquisition and manipulation required for performance analysis, 
modeling and optimization of codes on existing and emerging architectures. 
In this paper we introduce an initial implementation of the third component, 
a system for visualizing roofline charts and managing roofline performance
analysis data. We discuss the implementation and rationale for the
integration of the roofline visualization system into the Eclipse
IDE. An overview of our continuing efforts and goals in the development
of this project is provided.
\end{abstract}

\section{Introduction}

The Roofline model~\cite{Roofline2009} enables programmers to visualize the performance
potential of algorithms by introducing a simple 
way to quantify the computations' locality and parallelism and present them in the context of
a given architecture's capabilities. At present Roofline models are typically laboriously 
created through (1) collection of hardware performance data, e.g., with micro benchmarks; 
(2) manual code analysis to determine the arithmetic intensity of the algorithm(s) being studies; and
(3) visualizing both the architectural rooflines and the kernel's expected performance under 
different optimization assumptions. Automating most of this process is the goal of the 
Roofline Toolkit Project. The development of portable microbenchmarks that automate the first 
step is discussed in~\cite{RooflineLBL14}. In this paper we introduce the data representation
and visualization infrastructure required to automate the third step. We also discuss ongoing
work on partially automating the generation of performance models in the second step.

\subsection{Roofline Analysis}

For any machine model, we can evaluate the upper bound on performance by using the roofline 
model introduced by Williams et al~\cite{Roofline2009}. 
Given the arithmetic intensity of an algorithm, the roofline model 
defines an upper limit on kernel performance $P_k$ with the equation, $P_k = min{P_f , BA_i}$ 
where $P_f$ is the peak hardware floating-point performance, $B$ is peak bandwidth, 
and $A_i$ is the arithmetic intensity, typically expressed as the ratio of floating-point
operations to bytes transferred to/from memory.

The Roofline model and its extensions (e.g., for energy~\cite{Vuduc13}) provide a compact
representation of the \emph{architectural} capabilities as a context that enables 
visualization of a kernel's current and \emph{potential} performance within 
the algorithm and architectural constraints.


\subsection{Eclipse}

Eclipse~\cite{Eclipse} is a popular software platform with support for customized
IDE functionality. Its default set of plugins is designed for Java
development, but the Eclipse community has provided support for other
languages such as C/C++ and Fortran. Support for high performance
computing has also been provided via the Parallel Tools Platform (PTP)~\cite{PTP}. Two distinct advantages of the Eclipse platform are its portability
and extensibility. The former is provided largely by Eclipse\textquoteright{}s
Java-based implementation, which means it can be run consistently
on Windows, Macintosh and many Unix based OSes. Because Eclipse is
open source, users are free to modify and extend its functionality
as they see fit.

The value added by the Eclipse platform, both in terms of features
available to end users of the Roofline visualization framework and
APIs useful in the development of the framework made it an appealing
environment for integration of roofline visualization functionality.

\section{Visualization Implementation}

The initial roofline visualizations were implemented using general
purpose scientific charting tools such as GnuPlot. This was
adequate for developing and testing the roofline system and and for
the performance analysis activities of experts. However it was determined
early on that general adoption of the roofline system would benefit
from a simpler automated means of visualizing the performance data.
Furthermore, given the intended major use case of comparing the performance
of multiple applications or application routines to establish performant
behavior with respect to the roofline model a means of rapidly and
easily managing this process would be of benefit even to experts with
other visualization techniques at their disposal.

We implemented the charting system using JavaFX~\cite{JavaFX}. The new graphing
functionality provided in the JavaFX API allow reasonably sophisticated
visualizations without relying on external libraries so long as a
relatively recent version of Java is available. The specific requirements
of the roofline chart include a log2 scale axis, necessary for the
charted metrics to exhibit a parallel spatial relationship which simplifies
comparison between metrics. This had to be implemented manually based
on JavaFX's natural log scale axis. 

The data for visualizing Roofline architectural profiles is generated by a collection of portable 
micro benchmarks~\cite{RooflineLBL14}.
Currently the roofline visualization supports a single set of roofline
data. Multiple roofline datasets may be loaded simultaneously, either
from the local file-system or from an online repository, to allow
rapid switching between the data being visualized. The intersection
points and the inflection points on the roofline chart may be selected
to display the specific recorded metric values.

\begin{figure}
\centering
\begin{tabular}[t]{p{\textwidth}}
\parbox{1\textwidth}{\centering
\includegraphics[width=.65\textwidth]{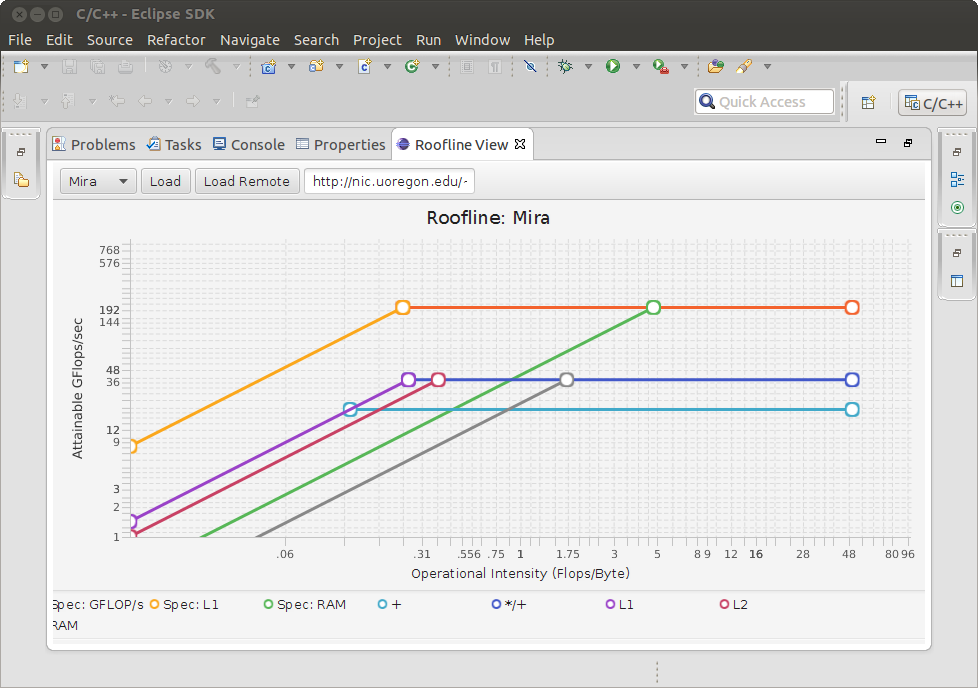}}\\
\subcaption{Roofline plot for Mira, an IBM Blue Gene/Q supercomputer. Nodes are 16-core PowerPC A2 (1.6 GHz) processors with 16GB.}\\
\parbox{1\textwidth}{\centering
\includegraphics[width=.65\textwidth]{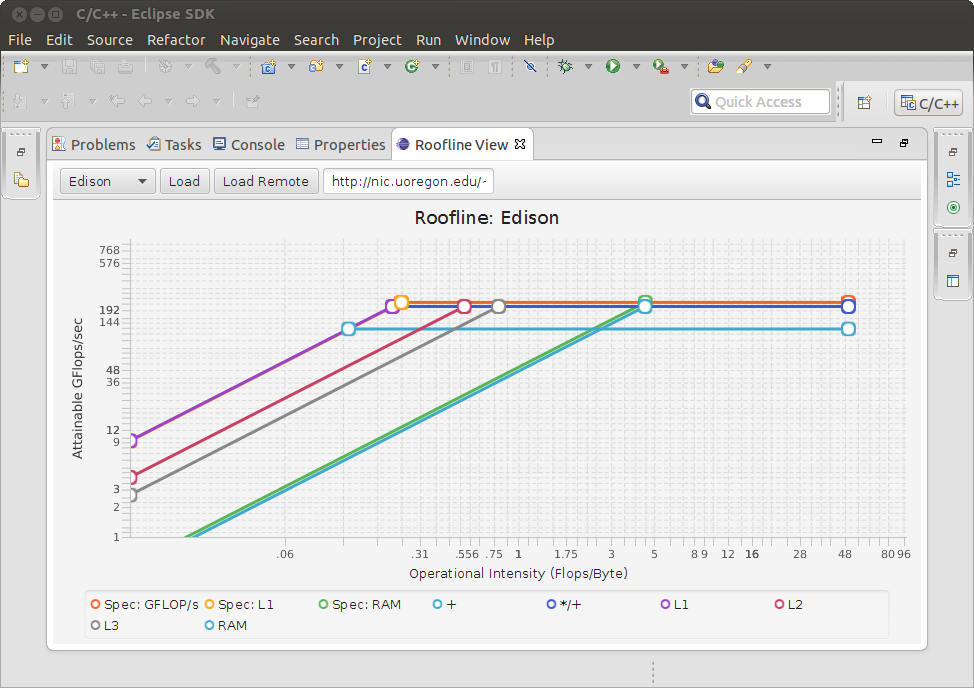}}\\
\subcaption{Roofline plot for Edison, a Cray XC30 supercomputer. Nodes are 12-core Intel "Ivy Bridge" processors (2.4 GHz) with 64 GB memory.}\\
\parbox{1\textwidth}{\centering
\includegraphics[width=.65\textwidth]{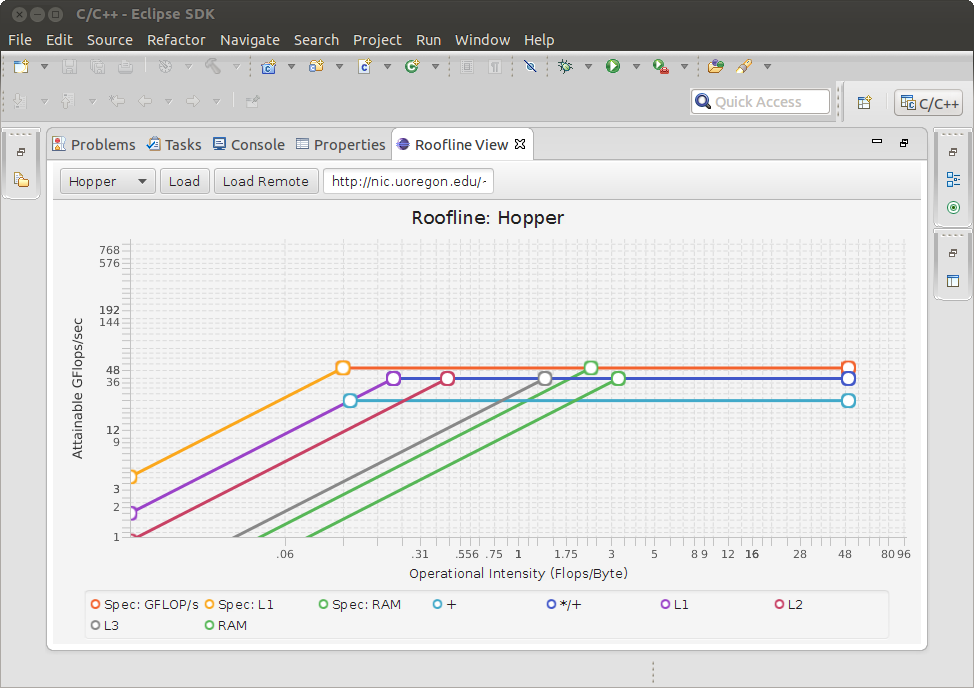}}\\
\subcaption{Roofline plot for Hopper, a Cray XE6 supercomputer. Nodes are two twelve-core AMD 'MagnyCours' (2.1-GHz) processors with 32 GB memory.}
\end{tabular}
\caption{Automatically generated roofline chart visualization. The top GFLOP/s rate  for the included benchmark data is indicated with an orange line and markers. In some cases lines corresponding to different hardware components overlap.}
\label{fig:example}
\end{figure}

Figure~\ref{fig:example} shows the rooflines generated 
for three architectures. Different rooflines reflect peak capabilities of 
different hardware components with respect to the operational intensity 
(x-axis) of the computation. For example, we can see that on Hopper,
a kernel whose operational intensity is 2 Flops/Byte can achieve 
near peak performance if it has relatively good L2 locality (but perhaps not L1) and 
takes advantage of fused multiply-add instructions. On the other hand, 
on Mira, the same kernel with
2 Flops/Byte operational intensity and an implementation with good L2 locality
and fused add-multiply use would achieve only a small fraction of
the potential peak (dark blue line vs orange horizontal line) unless it
also has good L1 locality.

\section{Data Management}

There are two primary elements to roofline charting data. The systems
where the roofline is being modeled are typified by the benchmarked
upper bounds on memory throughput and computational intensity. These
values and metric names are stored as name value pairs. The applications
being examined with respect to the system roofline are similarly stored
as simple pairs of recorded performance values and the name of the
application or subroutine. The simplicity of these data accommodate
a wide range of data presentation options. We have selected JSON~\cite{JSON} for
roofline data storage because it is simple to work with and there
is strong support for it in Java and Python. 

In addition to the core metrics of roofline system and application
performance data, the data format must accommodate a flexible system
for storing metadata for systems and experimental trials. This is
necessary to establish the provenance of collected data, to avoid
of duplicate trials and to allow searching and comparison of task
specific data from what may be a very large general collection of
system models and application trials. A robust medatadata system will
also support more advanced analytical features as those are provided
in future development.

The nature of roofline analysis lends itself to central, publicly
available data repositories. Because the system benchmarks are useful
to all developers working in the same environment it make sense to
make these a common resource. Publicly available application performance
data facilitates collaboration on performance tuning within a project
but also provides useful reference data for other users and developers
on the same system. Because of this the roofline visualization system
supports accessing roofline data from a remote repository. A preliminary
roofline data library is being assembled, hosted by the University
of Oregon.

\section{Future Work}

The roofline visualization system remains under heavy development
and there are a number of features we anticipate adding as the project
proceeds. These will dovetail with the continuing development of the
roofline data collection and analytical utilities also under development.

Improved and expanded visualization options are a fundamental component
of this undertaking. Comparison between systems and between multiple
sets of application trial data within a single chart will be useful
in performance engineering operations that incorporate roofline data.
This ties directly into the necessity for expanded and improved roofline
data search and storage capabilities. 

It is also our goal to increase the level of integration between the
Eclipse framework and the visualization system. In particular, for
developers working on projects in an Eclipse workspace, we would like
to allow direct navigation between source elements and the associated
roofline visualization. Using the Eclipse UI to control roofline data
collection and management is a feature which will be developed as
stable command line based roofline analytics tools become available.

We will also integrate the computation and visualization of arithmetic intensity 
(and other emerging algorithmic metrics) into the Eclipse environment,
so that users can easily visualize the current and potential performance of 
selected computations as they are developing them. To accomplish this
we will implement two main components of the Roofline Toolkit -- developer-aided
static model generation and empirical performance data integration. 
For static model generation, we will adapt the approaches used by source-based 
tools such as 
PBound~\cite{Nar:HPDC:2010} and binary analyzers such as MAQAO~\cite{MAQAO}.
The static approach allows the generation of more abstract models, which 
can then be modified by the user to reflect planned or necessary optimizations, 
with the resulting performance visualized in the context of architectural rooflines. 
Current performance stored in TAUdb databases will also be plotted in the 
new interface. Because while developing the software users focus on specific 
code portions, this will require automated extraction and agglomeration of 
performance data corresponding only to the selected code segments; in addition to 
GFLOPS/byte achieved, if more detailed hardware counter metrics are available, they 
will also be included in a separate detailed view. 

\section{Conclusion}

As the systems for conducting roofline analysis on high performance
computing systems improve and become more widely available we anticipate
a significant quantity of roofline data becoming publicly available
to support the performance engineering activities of software engineers.
The ability to easily conduct roofline analysis on local systems will
likewise improve as the project progresses. The ability to visualize
and make immediate use of this approach to performance analysis will
be simplified and streamlined by this visualization system.

\bibliographystyle{abbrv}
\bibliography{roofline}
\end{document}